\documentclass[published]{JHEP3}
\JHEP{00(2009)000}

\usepackage{epsfig,multicol,bbm}

\usepackage{amsmath}

\usepackage{amssymb}

\def\beq{\begin{equation}}
\def\eeq{\end{equation}}
\def\be{\begin{equation}}
\def\ee{\end{equation}}

\def\bea{\begin{eqnarray}}
\def\eea{\end{eqnarray}}

\title{Spectrum of Spin $1$ Dirac Operators on the Fuzzy $2$-Sphere}

\author{ Sanatan Digal\\ The Institute of Mathematical Sciences, CIT
Campus, Taramani, Chennai 600 113, India \\ E-mail:
\email{digal@imsc.res.in}}

\author{Pramod Padmanabhan \\ Department of Physics, Syracuse University, Syracuse, NY
13244-1130, USA \\ The Institute of Mathematical Sciences, CIT
Campus, Taramani, Chennai 600 113, India  \\ E-mail:
\email{ppadmana@syr.edu}}

\received{\today} \accepted{\today}

\abstract
{We numerically find out the spectrum of the $3$ spin $1$ Dirac
operators found in~\cite{ApbPP}. We give an analytic and numerical
proof that they are unitarily inequivalent. Since these operators
come paired with an anticommuting chirality operator, we find their
spectrums to resemble those of fermions with positive and negative
eigenvalues along with a number of zero modes. We give a method to count 
the number of zero modes which can be extended to higher spins on $S_F^2$.
An universal relation between the energy eigenvalues of the spin 1 Dirac 
operator and their multiplicities is found. This helps us predict the energy eigenvalues
for an arbitrarily large cut-off $L$, a problem which is computationally difficult
to handle.}

\keywords{Non-Commutative Geometry, Matrix Models, Field Theories in Lower Dimensions}

\begin{document}

\section{Introduction}
 Fuzzy spaces provide an attractive, alternative way to discretize
spacetime and thus help regularize field theories on such spaces.
Its interesting feature is the fact that this method preserves the
symmetries of the continuum spacetimes even at the discrete level
which are broken in lattice regularization techniques.

 Such fuzzy spaces are studied using the noncommuting algebra of
functions defined on them. One such widely studied model is the
fuzzy sphere $S_F^2$~\cite{Madore, Fuzzybook}. Several works on numerical
simulations of scalar fields and gauge fields on $S_F^2$ have 
been done.~\cite{Mp, Mp2, CrdSdTrg1,CrdSdTrg2, FgfXmDoc, DocBy}.

There have also been
interesting attempts to extend physics to other such fuzzy spaces
with higher genuses~\cite{Torus} and more exotic looking
surfaces~\cite{Cylinder,Beyondfuzzy}. But here we will consider only
$S_F^2$.

 In Connes' approach to noncommutative geometry~\cite{Connes},
the Dirac operator gains fundamental significance as part of the
spectral triple in formulating the spectral action principle. Such a
spectral action has been considered recently~\cite{WangWang} for the
Dirac operator of a spin $\frac{1}{2}$ particle. The operator used
by them corresponds to the one constructed in~\cite{Watamura}. In
this work we will consider Dirac operators constructed with the help
of Ginsparg-Wilson(GW) algebras. This approach provides an elegant
way to extend the construction of Dirac operators to all spins as
studied in~\cite{ApbPP}, where it was also found that several such
Dirac operators exist for the case of each spin $j$.

 We consider the spin $1$ case and numerically compute the spectrum
of the 3 Dirac operators. We also analytically compute the traces of
these $3$ operators and find that two of them have non-zero trace
showing the existence of unpaired eigenstates(zero modes). One of
them is traceless. Rather surprisingly, the spectrums are found to
be different, both numerically and, analytically from the trace
formulas,  thereby establishing the unitary inequivalence of the
three operators. The spectrum of the traceless Dirac operator is
studied in detail. The eigenvalues are plotted as a function of their degeneracy.
This is done for values of the cut-off, $L$ for which we could compute
the eigenvalues numerically.
We then fit this data with curves for each $L$ and find that a quadratic function
fits it. This fit extends rather amazingly to all other values of $L$ once we 
change the parameters in the fit which are functions of $L$. Thus this
gives an universal relation between the energy and their degeneracy and helps
predict the eigenvalues for any large value of $L$.

 The paper is organized as follows. Section 2 briefly describes the
noncommutative algebra on $S_F^2$. We then recall the construction
of Dirac operators using GW algebras in section 3. Here we also
write down the $3$ spin $1$ Dirac operators which we will work with. The trace of
these operators are found analytically. Their inequivalence provides a simple
proof for the unitary inequivalence of the three operators. In section 4 we recall
the spectrum of the spin $\frac{1}{2}$ Dirac operator and we study the properties
of the spectrum of the spin 1 Dirac operator. In particular we provide an elegant
way to count the number of zero modes of the traceless spin 1 Dirac operator for
each cut-off $L$. The numerical results are presented in section 5. We conclude in
section 6 with a few remarks and further speculations.

\section{Geometry of $S_F^2$}
 The algebra for the fuzzy sphere is
characterized by a cut-off angular momentum $L$ and is the full
matrix algebra $Mat(2L+1)\equiv M_{2L+1}$ of $(2L+1)\times (2L+1)$
matrices. They can be generated by the $(2L+1)$-dimensional
irreducible representation (IRR) of $SU(2)$ with the standard
angular momentum basis. The latter is represented by the angular
momenta $L^L_i$ acting on the left on $Mat(2L+1)$: If $\alpha\in
Mat(2L+1)$, \beq{L_{i}^{L}\alpha=L_{i}\alpha}\eeq
\beq{[L_{i}^{L},L_{j}^{L}]=i\epsilon_{ijk}L_{k}^{L}}\eeq
\beq{(L_{i}^{L})^2=L(L+1)\mathbf{1}}\eeq where $L_i$ are the
standard angular momentum matrices for angular momentum $L$.

 We can also define right angular momenta $L_i^R$:
\beq{L_{i}^{R}\alpha=\alpha L_{i}, \alpha\in M_{2L+1}}\eeq
\beq\label{commu}{[L_{i}^{R},L_{j}^{R}]=-i\epsilon_{ijk}L_{k}^R}\eeq
\beq{(L_{i}^{R})^2=L(L+1)\mathbf{1}}\eeq We also have
\beq{[L_i^L,L_j^R]=0.}\eeq

 The operator $\mathcal{L}_i=L_i^L-L_i^R$ is the fuzzy version of orbital
angular momentum. They satisfy the $SU(2)$ angular momentum algebra
\beq{[\mathcal{L}_i,\mathcal{L}_j]=i\epsilon_{ijk}\mathcal{L}_k}\eeq

 In the continuum, $S^2$ can be described by the unit vector
$\hat{x}\in S^2$, where $\hat{x}.\hat{x}=1$. Its analogue on $S_F^2$
is $\frac{L_i^L}{L}$ or $\frac{L_I^R}{L}$ such that
\beq{\lim_{L\rightarrow\infty}\frac{L_i^{L,R}}{L}=\hat{x}_i.}\eeq
This shows that $L_i^{L,R}$ do not have continuum limits. But
$\mathcal{L}_i=L_i^L-L_i^R$ does and becomes the orbital angular
momentum as $L\rightarrow\infty$:
\beq{\lim_{L\rightarrow\infty}L_i^L-L_i^R=-i(\overrightarrow{r}\wedge\overrightarrow{\nabla})_i
.}\eeq

\section{Construction of the Dirac Operators}
 In algebraic terms, the GW algebra $\mathcal{A}$ is the unital
$\ast$ algebra over $\mathbf{C}$ ,generated by two $\ast$-invariant
involutions $\Gamma, \Gamma'$.
 \beq \label{GW}\mathcal{A}=\{\Gamma,\Gamma'\ :\Gamma^2=\Gamma'^2=1\
,\Gamma^*=\Gamma\ ,\Gamma'^*=\Gamma'\}\eeq

 In any $\ast$ -representation on a Hilbert space,
$\ast$ becomes the adjoint $\dag$.

 Consider the following two elements constructed out of $\Gamma,
\Gamma'$: \beq {\Gamma_1=\frac{1}{2}(\Gamma+\Gamma'),}\eeq
\beq{\Gamma_2=\frac{1}{2}(\Gamma-\Gamma').}\eeq It follows from
Eq.(\ref{GW}) that $\{\Gamma_1,\Gamma_2\}=0$. This suggests that for
suitable choices of $\Gamma$, $\Gamma '$, one of these operators may
serve as the Dirac operator and the other as the chirality operator
provided they have the right continuum limits after suitable
scaling.

 For the spin $1$ case the combination which leads to the desired
Dirac and chirality operators were found in~\cite{ApbPP} and they
are \beq\label{D1}{ D_1 =
L\left(\frac{\Gamma_{L+1}^L-\Gamma_{L-1}^R}{2}\right),}\eeq
\beq\label{D2}{ D_2 =
L\left(\frac{\Gamma_{L-1}^L-\Gamma_{L+1}^R}{2}\right)}\eeq and
\beq\label{D3}{ D_3 =
L\left(\frac{\Gamma_{L}^L-\Gamma_{L}^R}{2}\right).}\eeq with
\beq\label{C1}{\gamma_1 =
\left(\frac{\Gamma_{L+1}^L+\Gamma_{L-1}^R}{2}\right),}\eeq
\beq\label{C2}{\gamma_2 =
\left(\frac{\Gamma_{L-1}^L+\Gamma_{L+1}^R}{2}\right)}\eeq and
\beq\label{C3}{\gamma_3 =
\left(\frac{\Gamma_{L}^L+\Gamma_{L}^R}{2}\right)}\eeq as their
corresponding chirality operators. In the above equations
\beq\label{Gl+1l}{\Gamma_{L+1}^{L}=\frac{2(\vec{\Sigma}.\vec{L}^L+L+1)(\vec{\Sigma}.\vec{L}^L+1)-(L+1)(2L+1)}{(L+1)(2L+1)},}\eeq
\beq\label{Gl+1r}{\Gamma_{L+1}^{R}=\frac{2(-\vec{\Sigma}.\vec{L}^R+L+1)(-\vec{\Sigma}.\vec{L}^R+1)-(L+1)(2L+1)}{(L+1)(2L+1)},}\eeq
\beq\label{Gl-1l}{\Gamma_{L-1}^{L}=\frac{2(\vec{\Sigma}.\vec{L}^L-L)(\vec{\Sigma}.\vec{L}^L+1)-L(2L+1)}{L(2L+1)},}\eeq
\beq\label{Gl-1r}{\Gamma_{L-1}^{R}=\frac{2(\vec{\Sigma}.\vec{L}^R+L)(\vec{\Sigma}.\vec{L}^R-1)-L(2L+1)}{L(2L+1)},}\eeq
\beq\label{Gll}{\Gamma^L_L=\frac{-2(\vec{\Sigma}.\vec{L}^L-L)(\vec{\Sigma}.\vec{L}^L+L+1)-L(L+1)}{L(L+1)},}\eeq
and
\beq\label{Glr}{\Gamma^R_L=\frac{2(\vec{\Sigma}.\vec{L}^R+L)(-\vec{\Sigma}.\vec{L}^R+L+1)-L(L+1)}{L(L+1)}.}\eeq
The operators in Eq.(\ref{Gl+1l})-Eq.(\ref{Glr}) are generators of
GW algebras and are obtained from left and right projectors to
eigenspaces of the total angular momentum, $\vec{L}+\vec{\Sigma}$,
where $\vec{\Sigma}$ are the matrices representing the spin $1$
representation of $SU(2)$.

 The continuum limits of Eq.(\ref{D1})-Eq.(\ref{D3}) are
\beq\label{CD1}{D_{1}=(\vec{\Sigma}.\vec{\mathcal{L}}-(\vec{\Sigma}.\hat{x})^{2}+2)+2(\vec{\Sigma}.\hat{x})+\{\vec{\Sigma}.\vec{\mathcal{L}},\vec{\Sigma}.\hat{x}\},}\eeq
\beq\label{CD2}{D_{2}=(\vec{\Sigma}.\vec{\mathcal{L}}-(\vec{\Sigma}.\hat{x})^{2}+2)-2(\vec{\Sigma}.\hat{x})-\{\vec{\Sigma}.\vec{\mathcal{L}},\vec{\Sigma}.\hat{x}\}}\eeq
and
\beq\label{CD3}{D_{3}=\vec{\Sigma}.\vec{\mathcal{L}}-(\vec{\Sigma}.\hat{x})^{2}+2.}\eeq

 The corresponding chirality operators in the continuum are \beq\label{CC1}{\gamma_{1}=(\vec{\Sigma}.\hat{x})^{2}+(\vec{\Sigma}.\hat{x})-1,}\eeq
\beq\label{CC2}{\gamma_{2}=(\vec{\Sigma}.\hat{x})^{2}-(\vec{\Sigma}.\hat{x})-1}\eeq
and \beq\label{CC3}{\gamma_{3}=1-2(\vec{\Sigma}.\hat{x})^{2}}\eeq
respectively.

\subsection*{The trace of the Dirac operators}
 The trace of the Dirac operators in Eq.(\ref{D1})-Eq.(\ref{D3}) can
be computed analytically by using the formula \beq{tr(A\otimes B)
=tr(A).tr(B)}\eeq where $A$ and $B$ are square matrices. Since the
Dirac operators we construct act on $Mat(2L+1)\otimes\mathbb{C}^3$,
they are of the form of tensor products and hence we can apply this
formula to analytically compute their traces.

 The trace is a rotationally invariant object leading to
\beq{tr((L_1^L)^2)=tr((L_2^L)^2)=tr((L_3^L)^2)}\eeq and
\beq{tr(\Sigma_1^2)=tr(\Sigma_2^2)=tr(\Sigma_3^2)=2.}\eeq The above
equations hold due to the fact that the three generators of any
representation of the $SU(2)$ algebra have the same trace because of
rotational invariance.

 The trace of $(L_i^L)^2$ varies according to whether $L$ is integer or
half-integer. When $L$ is an integer \beq{tr((L_i^L)^2) =
\frac{1}{3}L(L+1)(2L+1)^2}\eeq and when $L$ is an half-integer
\beq{tr((L_i^L)^2) = \frac{1}{3}L(L+1)(L+2)(2L+1).}\eeq The same formulas
hold when the left operators in the above equations are replaced by
right operators. It is simple to see that $\vec{\Sigma}.\vec{L}^L$
and $\vec{\Sigma}.\vec{L}^R$ are traceless. Using these identities
we write down the traces of our $3$ Dirac operators in
Table(\ref{sometable})

\begin{table}
\begin{center}
\begin{tabular}{|c|c|c|}
\hline
$\textrm{Dirac Operator}$ & $L \in \mathbb{Z}$ & $L \in \frac{\mathbb{Z}}{2}$ \\
\hline \hline $D_1$ & $4L(2L+1)$ & $2L(5L+1)$ \\ $D_2$ & $-4L(2L+1)$
& $-2L(5L+1)$ \\ $D_3$ & $0$ & $0$ \\ \hline \end{tabular}
\end{center}
\caption{Traces of the $3$ Dirac Operators} \label{sometable}
\end{table}

The trace of the Dirac operator is the sum of its eigenvalues. The
availability of these exact trace formulas are helpful in verifying
the spectrum of these operators found numerically.

 The operators $D_1$ and $D_2$ have non-zero trace implying the
existence of unpaired eigenstates or zero modes.

To check the unitary equivalence of the 3 Dirac operators, it is a necessary,
though not sufficient condition that the traces of the 3 operators be the same.
Since the trace formulas show the traces are not the same, they provide an
analytic proof for the unitary inequivalence of the 3 Dirac operators confirming
numerical results.

\section{Analytic results of the spectrums of the spin $\frac{1}{2}$ and spin 1 Dirac operators} 
 The spectrum of the spin $\frac{1}{2}$ Dirac operator can be found analytically~\cite{Fuzzybook}.  In the GW approach to constructing the Dirac operator, the spin $\frac{1}{2}$ system has the same spectrum both in the continuum and the fuzzy level. To illustrate the method of finding the spectrum, we consider the spin $\frac{1}{2}$ Dirac operator in the continuum: \beq\label{Dhalf}{D_{\frac{1}{2}} = \vec{\sigma}.\vec{\mathcal{L}}+1.}\eeq In the above equation $\vec{\mathcal{L}}$ is the orbital angular momentum got by taking the continuum limit of $\vec{L^L}-\vec{L^R}$. $\vec{\sigma}$ are the spin $\frac{1}{2}$ Pauli matrices. The total angular momentum $\vec{J}$ given by $$\vec{J}=\frac{\vec{\sigma}}{2}+\vec{\mathcal{L}}$$ commutes with the Dirac operator. We can use its eigenvalues to label the eigenstates of the Dirac operator. For  given cut-off angular momentum $L$, the spectrum of the orbital angular momentum is given by \beq\label{specL}{\vec{\mathcal{L}} \in \{0,1,\cdots , 2L\}.}\eeq Given this we can find the spectrum of the total angular momentum $\vec{J}$ to be \beq{\vec{J}\in \{\frac{1}{2}, \frac{3}{2}, \cdots , 2L-\frac{1}{2}, 2L+\frac{1}{2} \}. }\eeq Each value of the total angular momentum $\vec{J}$ can be got from two different orbital angular momentum except the top mode whose $\vec{J}$ value is $2L+\frac{1}{2}$. From this we can count the total number of eigenvalues for a given cut-off $L$ with the help of the following sum: \beq{\sum_{j=\frac{1}{2}}^{j=2L-\frac{1}{2}} 2(2j) + 2L+\frac{1}{2} = 2(2L+1)^2.}\eeq 
 
 The spectrum of the Dirac operator in Eq.(\ref{Dhalf}) can be got by noting that this operator can be written as \beq{D_{\frac{1}{2}}=\vec{J}^2-\vec{\mathcal{L}}^2+\frac{1}{4}.}\eeq As $\left[ \vec{J^2}, \vec{\mathcal{L}^2} \right]=0$, we can write the spectrum of $D_{\frac{1}{2}}$ as \beq{\textrm{Spectrum of}~ D_{\frac{1}{2}} = j(j+1) - l(l+1) +\frac{1}{4}.}\eeq As mentioned before each $j$ comes from two different $l$ values except the top mode. Thus we have for the spectrum of $D_{\frac{1}{2}}$: \beq{D_{\frac{1}{2}} \Bigg(\begin{split} & = j+\frac{1}{2}; ~~ \textrm{if}~ l= j-\frac{1} {2} \\ & = -j-\frac{1}{2}; ~~ \textrm{if}~ l =j+\frac{1}{2}. \end{split}\Bigg)}\eeq The spectrum has the chiral nature as expected. Note that there are no zero modes for the spin $\frac{1}{2}$ Dirac operator. The computation of the spectrum in the spin $\frac{1}{2}$ is easy due to the form of $D_{\frac{1}{2}}$ as given by Eq.(\ref{Dhalf}). This however is not true for the Dirac operator of the spin 1 case given by Eq.(\ref{CD3}). This is due to the presence of the term $\vec{\Sigma}.\hat{x}$ which does not commute with $\vec{\Sigma}.\vec{\mathcal{L}}$ making the analytic computation difficult. This is the reason why we take to numerical methods to achieve this. Nevertheless we can still get some vital information about the spectrum of the spin 1 Dirac operator by analytic methods.
 
 The total angular momentum $\vec{J}$ given by $$\vec{J}=\vec{\Sigma}+\vec{\mathcal{L}}$$ commutes with the Dirac operator in Eq.(\ref{CD3}) just as the corresponding total angular momentum does in the spin $\frac{1}{2}$ case. The spectrum of the orbital angular momentum $\vec{\mathcal{L}}$ is the same as in the spin $\frac{1}{2}$ case given by Eq.(\ref{specL}). The spectrum of $\vec{J}$ is now given by \beq{\vec{J} \in \{ 0, 1,2,\cdots , 2L-1, 2L, 2L+1\}.}\eeq In this case each value of $\vec{J}$ comes from three different orbital angular momenta namely $j-1$, $j$ and $j+1$ except three $j$ values. $j=0$ comes from only one state. $j=2L$ comes from 2 states and $j=2L+1$ comes from only one state. These are easy to check as they involve the simple angular momentum addition rules. With this information we can count the number of eigenvalues for each cut-off $L$ with the following sum:\beq{1+\sum_{j=1}^{j=2L-1} 3(2j+1) + 2(4L+1) + 2(2L+1)+1 = 3(2L+1)^2.}\eeq This is exactly the number of eigenvalues we expect from each cut-off $L$ for the spin 1 case as this is the size of the matrix for the Dirac operator for each $L$. These arguments can be easily extended to the Dirac operators of all spins but we will not do so here.
 
 \subsection{Number of positive eigenvalues and Zero modes for the spin 1 Dirac operator}
 Out of the three Dirac operators in the spin 1 case we will consider the traceless Dirac operator~(See Table \ref{sometable}). The trace equation gives us an easy and elegant way to count the number of different non-zero positive and negative eigenvalues as well as the number of zero modes for each cut-off angular momentum $L$.  
 
 The zero modes can be counted as follows: $j=0$ comes from just one orbital angular momentum state and so it cant result in a positive or negative eigenvalue of $D_3$ and hence it must only be 0 due to the traceless nature of the Dirac operator. This contributes 1 zero mode for each $L$. Similar argument holds for $j=2L+1$ which contributes $2(2L+1)+1$ zero modes for each $L$. For values of $j$ between 1 and $2L-1$ there is a contribution of $2j+1$ zero modes for each of the $j$ values. Summing all this we find that there are exactly $(2L+1)^2+2$ zero modes for each $L$. 
 
 In a similar way we can find the number of positive eigenvalues. When we do this we find there is a contribution of $2j+1$ eigenvalues for values of $j$ between 1 and $2L$. Summing these we get $4L^2+4L$. As the Dirac operator is traceless, the same argument holds for the negative eigenvalues giving a total of $4L^2+4L$ eigenvalues for each $L$. It is easy to see that the sum of the positive, negative eigenvalues and zero eigenvalues give $3(2L+1)^2$ as the total number of eigenvalues as expected for each cut-off $L$. 
 
 These arguments can again be easily extended to the spectrum of higher spin Dirac operators on $S_F^2$. It is also easy to see that there are no zero modes for half-integral spin systems on $S_F^2$ as none of the Dirac operators for half-integral spin systems are traceless. We will not discuss them any further in this work except for a few remarks in the end.
 
 Finally we count the number of different positive eigenvalues we expect to find for the spin 1 Dirac operator for each cut-off $L$. Since there are $2L+2$ values  the total angular momentum $j$ can take, out of which 2 of them can only contribute to the zero modes for each $L$, we can conclude that there are $2L$ different positive eigenvalues for each $L$. The degeneracies of each of them can easily be read off as $2j+1$ according to the corresponding value $j$ takes. 
 
 \subsection{Remarks on the other two Dirac operators for the spin 1 case}
 So far the arguments in this section were for the traceless Dirac operator in Eq.(\ref{CD3}). These arguments do not hold for the Dirac operators in Eq.(\ref{CD1}) and Eq.(\ref{CD2}) as they have positive and negative traces respectively. These are given in table \ref{sometable}. 
 
 Consider the Dirac operator with the positive trace whose continuum value is given by Eq.(\ref{CD1}). In this case too we have for the spectrum of the total angular momentum \beq{\textrm{Spec}~\vec{J}\in\{0,1,\cdots , 2L-1, 2L, 2L+1\}}\eeq as before. However in this case we cannot say that the states corresponding to $j=2L+1$ and $j=0$ correspond to zero modes. This is because of the non-zero trace. They now have some positive energy say $E_0$ and $E_{2L+1}$. We then have the following equation \beq{E_0 + (4L+3)E_{2L+1}= 4L(2L+1)}\eeq  for integral values of $L$ and \beq{E_0 + (4L+3)E_{2L+1}= 2L(5L+1)}\eeq for half-integral values of $L$. The $4L+3$ states with energy $E_{2L+1}$ correspond to unpaired eigenstates. If $|E_{2L+1}\rangle$ is the state with energy $E_{2L+1}$, then these states will be of the form $\gamma^{2k}|E_{2L+1}\rangle$ where $\gamma$ is the chirality operator given in Eq.(\ref{CC1}) and $k$ is an integer. In the spin $\frac{1}{2}$ case these states, with the corresponding chirality operator for the spin $\frac{1}{2}$ system, will equal $|E\rangle$ itself as $\gamma ^2=1$ for the spin $\frac{1}{2}$ case. It can be easily seen from Eq.(\ref{CC1}) that this is not true for the spin 1 case. So we get the possibility for a number of states with the same energy. With this note, we analyze only the spectrum of the traceless Dirac operator in what follows.

 Having studied the general nature of the spectrum for the spin 1 Dirac operator, we compute the eigenvalues numerically in the next section. In particular we will find a relation between the eigenvalue and its multiplicity for a given cut-off $L$. This is equivalent to finding the eigenvalues as a function of total angular momentum $j$ for each cut-off $L$.

\section{Numerical Results}
 We compute the eigenvalues of the the three Dirac operators in
Eq.(\ref{D1})-Eq.(\ref{D3}) numerically. The size of each of these
operators is $3N^2$ where $N=2L+1$ and $L$ is the cutoff. It is
clear from the dimensions of these matrices($\sim 9N^4$) that we
cannot go to arbitrarily large values of $N$.  Even for $L=22$, the
size of the matrix becomes $6075\times 6075$ which is difficult to
handle numerically within the resources available to us. For large values of $N$
number of computational steps increase which will lead to growth of
systematic error. However,
the patterns emerging from the spectrum we computed so far strongly
suggest what the behavior would be at higher values of $N$. This circumvents
computational problems and helps us predict the behavior as we go
close to the continuum. This is particularly important given the problems in handling very large matrices.

The nature of the spectrum was discussed in the previous section and we confirm those results numerically. 
 The spectrum of $D_3$ is similar to that of fermions with equal
number of positive and negative energy eigenvalues. This is a
reflection of the existence of the chirality operator given by
Eq.(\ref{C3}), which anticommutes with $D_3$. Apart from the
non-zero eigenvalues there also exist a number of zero modes. We find exactly $(2L+1)^2+2$ zero modes for each cut-off $L$ as we explained in the previous section. The number of positive eigenvalues is also as expected.
 
 We work with only the positive eigenvalues of the spectrum. As the operator is traceless we have the same pattern for the negative eigenvalues and so we do not use them to fit curves. Then we find the degeneracies of each of the positive eigenvalues. Note from the discussion in the earlier section, that there can only be odd degeneracies for our system as the total angular momentum $j$ takes integral values. The plot for the energy vs the degeneracies is shown in figure 1.  It shows  the data points for three different values of $N$ (namely $N=21$, $N=35$ and $N=45$) along with the best fit curves. 

\begin{figure}
\begin{center}
\mbox{
    \leavevmode
   { 
      \includegraphics[height=4.5in,width=2.5in, angle=270]{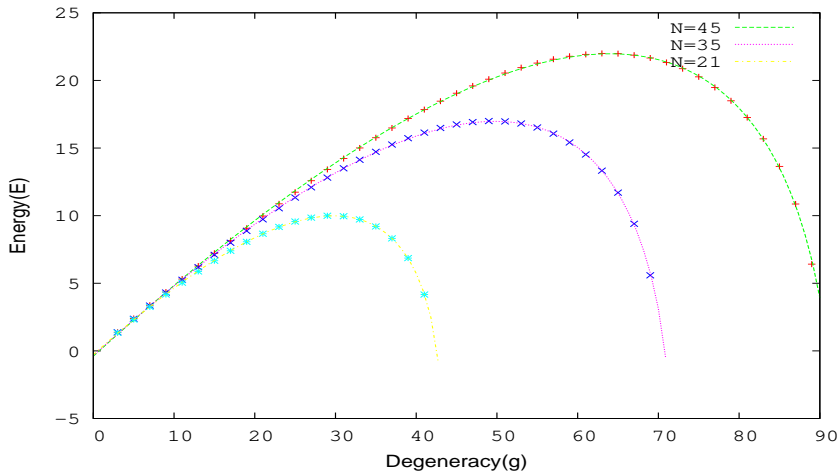} }}
 \end{center}
   \caption{Plot of the energy eigenvalues along with the best fit curves for $N=21$, $N=35$ and $N=45$.}
\end{figure}

 By inspection we found the curve has a mirror symmetry about some principal axis. 

 Next we try to find a universal curve that will fit the data(eigenvalues) of different cutoffs, just by changing the value of the cutoff. To this end we analyzed the data in a rotated frame in which the data was found to have reflection symmetry around the rotated y-axis. Given that the data for small $(E,g)$
is independent of the cutoff(this can be seen in figure 1 where for small values of $g$ the three sets of data points lie almost on top of each other on a straight line), we found a unique rotation angle to rotate all the results for different cut-off $L$. After observing the reflection symmetry of
$(E^\prime,g^\prime)$ we tried to fit the data with a polynomial with only even powers. To our surprise we found an excellent fit with just a
parabola for all different cutoff values. Higher powers in the function did not make any further improvement in the fitting. The parameters of
the parabola run with the cut off. We also find excellent fit for these parameters as a function of the cutoff $L$. 

 We now elaborate this method. The plot of $(E,g)$ is rotated to a new set of variables $(E',g')$. This set of points is then fitted with the curve \beq\label{energyprime}{E'=\alpha(g'+\eta)^2+\beta.}\eeq Here $\alpha$, $\beta$ and $\eta$ are expected to vary with the cut-off $L$.  The relation between $(E',g')$ and $(E,g)$ is given by \beq\label{transform}{\left(\begin{array}{c} E' \\ g' \end{array}\right) = \left(\begin{array}{cc} \cos{\theta} & \sin{\theta} \\ -\sin{\theta} & \cos{\theta} \end{array}\right) \left(\begin{array}{c} E \\ g\end{array}\right).}\eeq The angle $\theta = 2.26159 \textrm{radians}$. This angle is a constant for different values of $L$. This can be seen as a consequence of the data points lying on top of each other for small values of $g$ as seen in figure 1. Note that $E'$ and $g'$ are not energies and degeneracies respectively. We just need to use the transformation in Eq.(\ref{transform}) to get the relation between the energies and the degeneracies. 
 
 Our next task is to find $\alpha$, $\beta$ and $\eta$ as functions of $N=2L+1$. They are found by fitting the quadratic form (Eq.(\ref{energyprime})) to the rotated curves for different values of $N$. We find them to follow simple relations. These are shown in the figures 2. The exact functions we found were: \beq{\alpha = \frac{0.863569}{N^{0.930775}}-0.00141635,}\eeq \beq{\beta = -1.45123 N +0.333497}\eeq and \beq{\eta=-1.16288 N - 0.555529.}\eeq The numbers may look uninteresting but if we could fit these functions after we find these parameters for more values of $N$ we could converge onto some special numbers. We did not attempt this in this work. The relations are simple enough to imply something deeper in the spectrum. More exact numbers could help in the quest for an analytic solution of this problem.

\begin{figure}
\begin{center}
\mbox{ \leavevmode
{ \label{f:subfig-1}
\includegraphics[height=2.15in,width=1.75in,angle=270]{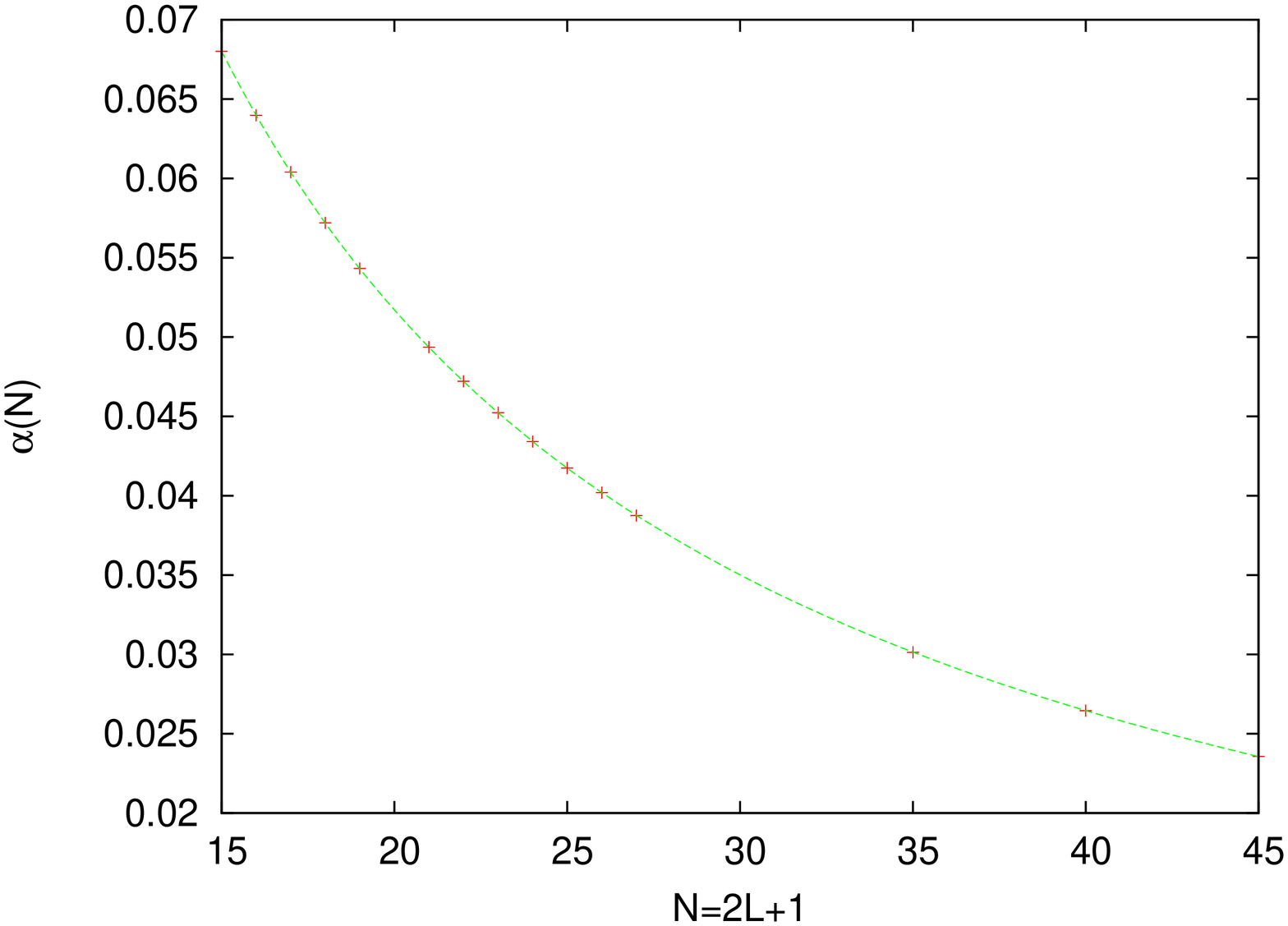} }
\leavevmode
{ \label{f:subfig-2}
\includegraphics[height=2.15in,width=1.75in, angle=270]{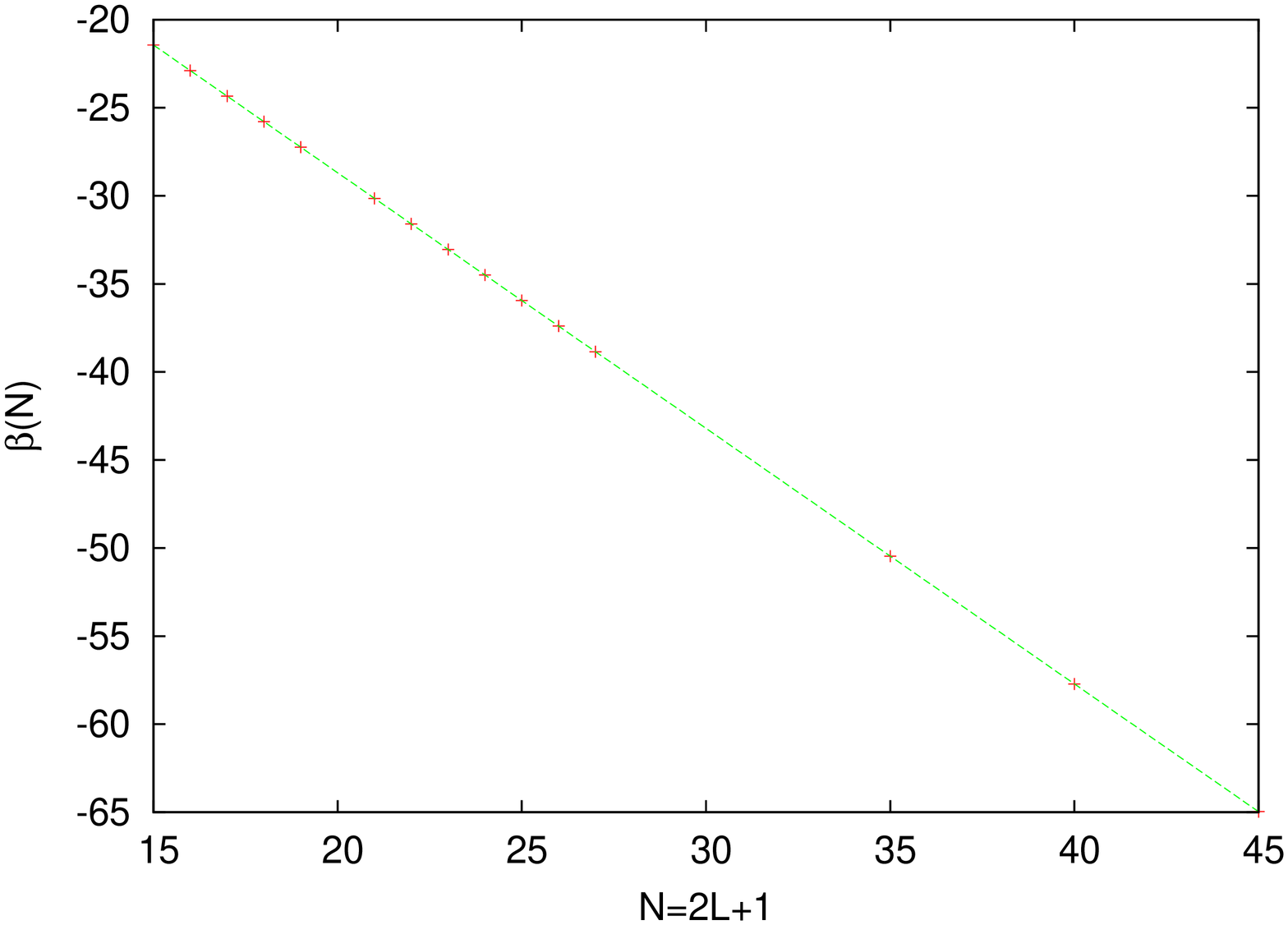} }
\leavevmode
{ \label{f:subfig-3}
\includegraphics[height=2.15in,width=1.75in, angle=270]{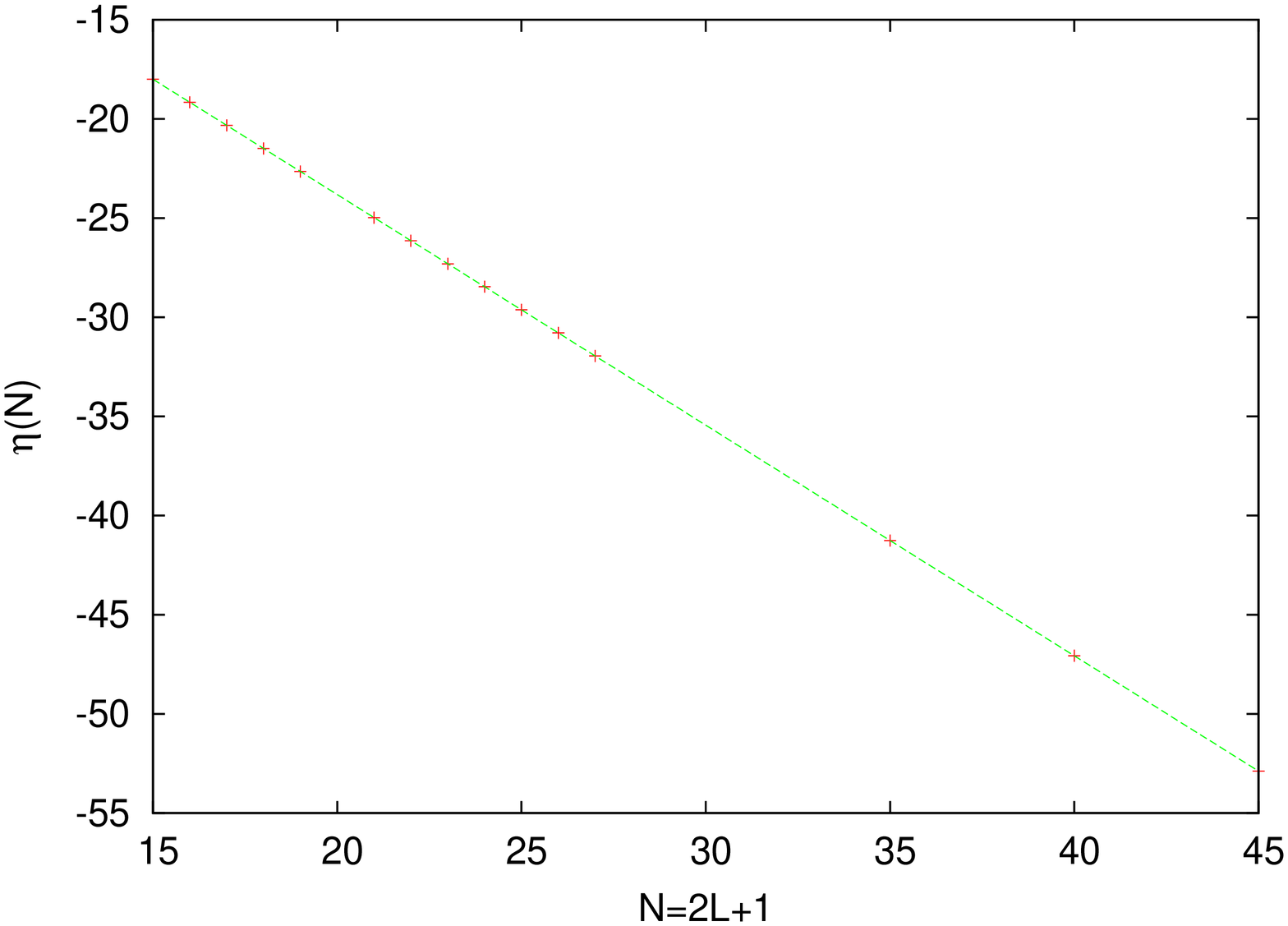} }}
\end{center}
\caption{Parameters $\alpha$, $\beta$ and $\eta$ as a function of $N$.}
\end{figure}
 
 We can now write down the exact relation between $E$ and $g$ based on our numerical fits: \beq\label{energy1}{ E = \frac{\sqrt{b(g)^2-4ac(g)}-b(g)}{2a}}\eeq where \beq{a = \alpha \cos ^2{\theta},}\eeq \beq{b(g)=2\alpha\cos{\theta} \left(\sin{\theta}g+\eta\right)+\sin{\theta}}\eeq and \beq{c(g)=\alpha\left(\sin{\theta}g+\eta\right)^2+\beta-\cos{\theta}g.}\eeq
 
 Having found this relation between $E$ and $g$, we can now find the eigenvalues for arbitrarily large values of $N$. If we diagonalize $D_3$ for such large values of $N$ it would take a lot of memory on the computer and is subject to a lot of numerical error. But we can get around this with our relation between $E$ and $g$. Figure 3 shows the eigenvalues for $N=60$. Note that though the curve looks continuous, we have seen in the previous section that degeneracies are allowed to take only odd integral values. The maximum degeneracy for a given $N$ is $2N-1$. Starting from 3 we can allow $g$ to vary till $2N-1$ through odd integers and find the corresponding eigenvalues using Eq.(\ref{energy1}). In an equivalent manner we can find the energy eigenvalues as a function of the total angular momentum $j$ by simply substituting $g=2j+1$ in Eq.(\ref{energy1}).  

\begin{figure}
\begin{center}
\mbox{
    \leavevmode
    { 
      \includegraphics[height=4.5in,width=2.5in,angle=270]{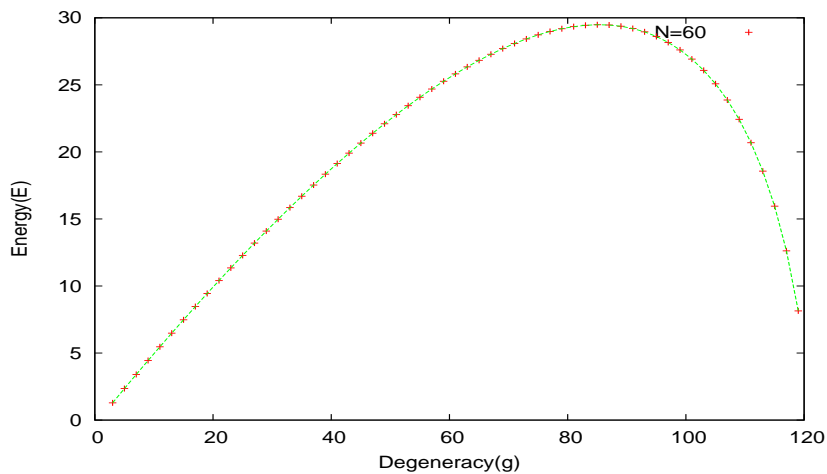} }}
\end{center}
\caption{Plot predicting the energy eigenvalues for $N=60$.}
\end{figure}

\section{Conclusions}
 The spectrums of the spin $1$ Dirac operators are found numerically.
The three operators do not have the same spectrum making them
unitarily inequivalent. This may have interesting consequences which
we plan to explore in the future. The fermionic character of the
spectrum is noteworthy as there exists no such higher dimensional
analog in the Minkowski case. We expect this behavior also for
higher spin Dirac operators on $S_F^2$ as they all come paired with
an anticommuting chirality operator.

The universal relation between the energy eigenvalues and their degeneracies
we find in Eq.(\ref{energy1}) is indeed remarkable as it circumvents what 
would have been an almost impossible computational problem involving large
matrices. This now allows us to find the eigenvalues for any arbitrarily large
cut-off $L$. The simple relation between the eigenvalues and their degeneracies seem
to suggest some connection with the underlying symmetry in the problem.

 Having obtained the spectrum for the spin 1 Dirac operator for arbitrarily large cut-off $L$,
we can go on to find the partition function for a system of particles occupying
these energy levels. Assuming fermionic statistics we can compute
several thermodynamic quantities for this system. 
Several interesting features arise and these are reported in~\cite{Future}.

 A quantum particle on the continuum sphere, $S^2$ has energy eigenvalues
given by $l(l+1)$. These are the eigenvalues of the Laplacian on the
sphere which is a second order differential operator. The
eigenvalues of the square of the continuum limit of the spin
$\frac{1}{2}$ Dirac operator on
$S_F^2$~\cite{Watamura,ApbPP,Fuzzybook,Presnajder} also gives a
spectrum similar to that of the standard Laplacian on $S^2$ apart
from a additional constant. This additional constant can be
interpreted as the scalar curvature according to the Lichnerowicz
formula for the square of a general Dirac operator. In the
Minkowskian case this is analogous to the square of the Dirac
operator giving the Laplacian on that space. This leads to each
component of the Dirac spinor satisfying the Klein-Gordon equation.
We can view the Laplacian of the standard sphere as an analog of the
Klein-Gordon equation on the sphere as this gives the $SU(2)$
covariant dispersion relation on $S^2$. Note that we can add
additional constants to this Laplacian as they are rotationally
invariant. This however is not true for any of the $3$ spin $1$
Dirac operators on $S_F^2$ as their continuum limits, given by
Eq.(\ref{CD1})-Eq.(\ref{CD3}), contain $\vec{\Sigma}.\hat{x}$ terms
which makes the square of these operators look complicated. (Note
that we do still get $l(l+1)$, but with additional terms containing
$\vec{\Sigma}.\hat{x}$ which makes the analytical computation of the
spectrum difficult.) Thus the spectrum of their squares are not the
standard one making the study of these deviations interesting as
there exist no counterparts on higher dimensional Minkowskian space.

 We are also computing the spectral action of these Dirac operators.
This will be compared with the spectral action of the spin
$\frac{1}{2}$ Dirac operator on the continuum sphere. This kind of
analysis was carried out recently~\cite{WangWang} where interesting
connections were made with cosmology. The results will be reported
in a future work~\cite{Future1}.

 It was found in~\cite{ApbPP}, that for a given chirality operator
there exist several different Dirac operators. However this was done
in the continuum limit and we have not found their fuzzy analogs. These will
most certainly not be unitarily equivalent. This seems to be a new property 
of spin systems on $S_F^2$ and they need to be studied further.

\section{Acknowledgements}
 We thank Prof.A.P.Balachandran, Prof.T.R.Govindrajan and Prof. Satyavani Vemparala
for useful discussions and references. We also thank Prof.Xavier
Martin for useful comments. 
 PP thanks Prof.T.R.Govindarajan for the hospitality at IMSc,
Chennai. This work was supported in part by DOE under the grant
number DE-FG02-85ER40231.

\bibliographystyle{apsrmp}
{99}

\end{document}